\documentstyle[twoside]{article}

\catcode`\@=11
\long\def\@makefntext#1{
\protect\noindent \hbox to 3.2pt {\hskip-.9pt  
$^{{\eightrm\@thefnmark}}$\hfil}#1\hfill}

\def\thefootnote{\fnsymbol{footnote}}
\def\@makefnmark{\hbox to 0pt{$^{\@thefnmark}$\hss}}
	
\def\ps@myheadings{\let\@mkboth\@gobbletwo
\def\@oddhead{\hbox{}
\rightmark\hfil\eightrm\thepage}   
\def\@oddfoot{}\def\@evenhead{\eightrm\thepage\hfil
\leftmark\hbox{}}\def\@evenfoot{}
\def\sectionmark##1{}\def\subsectionmark##1{}}

\oddsidemargin=\evensidemargin
\renewcommand{\thefootnote}{\fnsymbol{footnote}}

\newcounter{sectionc}\newcounter{subsectionc}\newcounter{subsubsectionc}
\renewcommand{\section}[1] {\vspace{12pt}\addtocounter{sectionc}{1} 
\setcounter{subsectionc}{0}\setcounter{subsubsectionc}{0}\noindent 
	{\tenbf\thesectionc. #1}\par\vspace{5pt}}
\renewcommand{\subsection}[1] {\vspace{12pt}\addtocounter{subsectionc}{1} 
	\setcounter{subsubsectionc}{0}\noindent 
	{\bf\thesectionc.\thesubsectionc. {\kern1pt \bfit #1}}\par\vspace{5pt}}
\renewcommand{\subsubsection}[1] {\vspace{12pt}\addtocounter{subsubsectionc}{1}
	\noindent{\tenrm\thesectionc.\thesubsectionc.\thesubsubsectionc.
	{\kern1pt \tenit #1}}\par\vspace{5pt}}
\newcommand{\nonumsection}[1] {\vspace{12pt}\noindent{\tenbf #1}
	\par\vspace{5pt}}

\newcommand{\textlineskip}{\baselineskip=13pt}
\newcommand{\smalllineskip}{\baselineskip=10pt}

\def\abstracts#1#2#3{{
	\centering{\begin{minipage}{4.5in}\baselineskip=10pt\footnotesize
	\parindent=0pt #1\par 
	\parindent=15pt #2\par
	\parindent=15pt #3
	\end{minipage}}\par}} 

\newcommand{\bibit}{\nineit}
\newcommand{\bibbf}{\ninebf}
\renewenvironment{thebibliography}[1]
	{\frenchspacing
	 \ninerm\baselineskip=11pt
	 \begin{list}{\arabic{enumi}.}
	{\usecounter{enumi}\setlength{\parsep}{0pt}
	 \setlength{\leftmargin 12.7pt}{\rightmargin 0pt}
	 \setlength{\itemsep}{0pt} \settowidth
	{\labelwidth}{#1.}\sloppy}}{\end{list}}

\newcounter{itemlistc}
\newcounter{romanlistc}
\newcounter{alphlistc}
\newcounter{arabiclistc}

\def\pmb#1{\setbox0=\hbox{#1}
	\kern-.025em\copy0\kern-\wd0
	\kern.05em\copy0\kern-\wd0
	\kern-.025em\raise.0433em\box0}

\def\fnt#1#2{\footnotetext{\kern-.3em
	{$^{\mbox{\scriptsize #1}}$}{#2}}}

\def\fpage#1{\begingroup
\voffset=.3in
\thispagestyle{empty}\begin{table}[b]\centerline{\footnotesize #1}
	\end{table}\endgroup}

\font\tenrm=cmr10
\font\tenit=cmti10 
\font\tenbf=cmbx10
\font\bfit=cmbxti10 at 10pt
\font\ninerm=cmr9
\font\nineit=cmti9
\font\ninebf=cmbx9
\font\eightrm=cmr8

\newcommand{\coma}[1]{\,#1}
\newcommand{\ecuacion}[2]{\begin{equation} #1 \coma{#2} \end{equation}}
\newcommand{\nueva}{\nonumber \\}
\newcommand{\ptc}[1]{\coma{#1} \nueva}
\newcommand{\ecuarreglo}[2]{\begin{eqnarray} #1 \coma{#2} \end{eqnarray}}	
\newcommand{\refeq}[1]{(\ref{#1})}
\newcommand{\letra}[1]{{\it #1}}
\newcommand{\revista}[5]{#1, {\bibit #2} {\bibbf #3}, (#4) #5}
\newcommand{\libro}[6]{#1, {\bibit #2}, #3 (#4, #6, #5)}
\newcommand{\citar}[1]{$^{#1}$}

\newcommand{\iz}{\left}
\newcommand{\de}{\right}
\newcommand{\espacio}{\qquad \qquad}
\newcommand{\ld}[1]{\lambda^{#1}}
\newcommand{\deriv}[1]{\partial_#1}
\newcommand{\lag}{{\cal L}}
\newcommand{\tr}{\mathop {\rm Tr}\nolimits}
\newcommand{\diag}{\mathop {\rm diag}}
\newcommand{\spart}{\hbox{$\not \! \partial$}}
\newcommand{\scova}{\hbox{$\not \! \! D$}}
\newcommand{\sgauge}{\hbox{$\not \! \! A$}}
\newcommand{\sgaugeB}{\hbox{$\not \! \! B$}}
\newcommand{\higgs}{\varphi}
\newcommand{\g}[1]{\gamma^#1}
\newcommand{\Amu}{A_\mu^{\ a}}
\newcommand{\eL}{$e_L$}
\newcommand{\eR}{$e_R$}
\newcommand{\eC}{$e_L^c$}
\newcommand{\ve}{$\nu_L$}
\newcommand{\uno}{{\bf 1}}
\newcommand{\bfsigma}{\hbox{{\boldmath $\sigma$}}}
\newcommand{\bfsgauge}{\hbox{$\bf \not \! \! A$}}
\newcommand{\uu}[1]{$U(#1)$}
\newcommand{\bfuu}[1]{{\boldmath $U$}{\bf (}$\bf #1${\bf )}}
\newcommand{\su}[1]{$SU(#1)$}
\newcommand{\bfsu}[1]{{\boldmath $SU$}{\bf (}$\bf #1${\bf )}}

\newcommand{\suu}{$SU(2) \otimes U(1)$}
\newcommand{\bfsuu}{\bfsu2{\boldmath $\, \otimes\,$}\bfuu1}

\newcommand{\susu}{$SU(2)_L \otimes SU(2)_R$}
\newcommand{\gsu}{$SU(2/1)$}
\newcommand{\gsuA}{$su(2/1)$}

\begin{document}

\normalsize\textlineskip
\thispagestyle{empty}
\setcounter{page}{1}

\vspace*{-2.5cm}\smalllineskip{\flushright
        {\footnotesize UCR-EF-98-3}\\}

\vspace*{0.88truein}

\fpage{1}
\centerline{\bf UNIFICATION OF \bfsuu\ USING A GENERALIZED}
\vspace*{0.035truein}
\centerline{\bf COVARIANT DERIVATIVE AND \bfuu3}
\vspace*{0.37truein}
\centerline{\footnotesize M. CHAVES AND H. MORALES}
\vspace*{0.015truein}
\centerline{\footnotesize\it Escuela de F\'{\i}sica, Universidad de Costa Rica,}
\baselineskip=10pt
\centerline{\footnotesize\it Ciudad Universitaria Rodrigo Facio, San Jos\'e, Costa Rica}

\vspace*{0.225truein}
\abstracts{
A generalization of the Yang-Mills covariant derivative, that uses both 
vector and scalar fields and transforms as a 4-vector contracted with Dirac 
matrices, is used to simplify the Glashow-Weinberg-Salam model. 
Since \su3\ assigns the wrong hypercharge to the Higgs boson, it 
is necessary to use a special representation of \uu3\ to obtain all the correct 
quantum numbers. A surplus gauge scalar boson emerges in the process, 
but it uncouples from all other particles.}{\vspace*{7pt} \parindent=0pt PACS numbers: 12.60.Jv 11.15.-q 12.15.-y 12.60.Fr}{}

\textheight=7.8truein
\setcounter{footnote}{0}
\renewcommand{\thefootnote}{\alph{footnote}}

\vspace*{1pt}\textlineskip
\section{Introduction}
\vspace*{-0.5pt}
\noindent
	In spite of its great successes, the Glashow-Weinberg-Salam (GWS) model, based on the 
group \suu, seems somewhat unsatisfactory, in that its lagrangian contains a number of 
arbitrary parameters and several terms that seem, at least at first sight, rather \letra{ad hoc}. There have 
been different types of ideas to improve this situation. The most popular one by far has been to 
use a large group of which \suu\ is just a small subgroup. Less popular has been the 
introduction of symmetries that do not enlarge the group too much, a step-at-a-time policy, so to 
speak. This has been done along three different lines, depending on what additional symmetry is 
imposed, be it the group \susu,\citar{1} a discrete symmetry that restricts the model's 
parameters and allows for the calculation of mass corrections to quarks,\citar{2} or \su3. Here we 
shall concentrate our efforts on this last idea, not following the usual Yang-Mills scheme,\citar{3} but 
instead considering a modification of it.

	It is interesting that \su3\ contains \suu\ as a subgroup and that it naturally 
assigns the correct hypercharge $Y$ and isospin $T_3$ quantum numbers to the electron and the 
neutrino if we place them in its fundamental representation so that they form the chiral triplet
\ecuacion{
\psi_L = \pmatrix{ \nu_L \cr e_L \cr e_L^c }
\label{chiral}
}.

	It is necessary to express the \eR\ degree of freedom using the antiparticle formulation \eC\ in order 
to obtain the hypercharge with the correct sign, as we shall soon explain. The main problem that 
presents itself immediately when one tries to represent the GWS model using \su3\ stems from 
the boson sector, because 8 vector fields are required to gauge the theory, one for every generator 
of the adjoint representation. This means that there would be four more boson vector fields than 
are experimentally observed.

	A different approach is based not on Lie but rather on graded groups\citar{4} and uses \gsu. 
The excessive number of vector bosons introduced by the gauging of \su3\ is reduced by 
assuming that some of them are scalar instead of vectorial. It is possible to use the Higgs bosons 
as gauge fields, thus adding to the logical simplicity of the model. Fermions are placed in the 
fundamental representation of \gsu\ using the nonchiral triplet
\ecuacion{
\psi = \pmatrix{ \nu_L \cr e_L \cr e_R }
\label{nonchiral}
}, 
that differs from the chiral one in the third component. The generators of the fundamental 
representations of \su3\ and \gsu\ differ in only one component of one generator. We are 
going to use the generators $T^a=\frac{1}{2} \ld{a}$, where the $\ld{a}$ are the usual Gell-Mann matrices, so that 
they are normalized according to
\ecuacion{
\widetilde{\tr}\, T^a T^b = \frac{1}{2} \delta_{ab}
\label{norma}
},
where we have put a tilde over the trace symbol to distinguish it from the traces over Dirac 
matrices that we shall begin to use next section. The two generators that differ are both diagonal: 
the last \su3\ Gell-Mann generator
\ecuacion{
\ld{8} = \frac{1}{\sqrt{3}} \diag (1,1,-2)
\label{gene8}
},
and its partner in \gsu
\ecuacion{
\ld{0} = \frac{1}{\sqrt{3}} \diag (1,1,2)
\label{gene0}
}.

	The isospins of \ve, \eL\ and \eR\ are $\frac{1}{2}$, $-\frac{1}{2}$ and $0$, and the hypercharges are $1$, $1$ and $2$, respectively, 
precisely as given by $\ld{0}$. The electric charges of these particles can be calculated from the Gell-Mann-Nishijima 
relation $Q=T_3-\frac{1}{2} Y$, that relates the charge $Q$ to the isospin $T_3$ and the 
hypercharge $Y$. From these considerations it is clear that $\ld{8}$ would assign a hypercharge $-2$ for 
\eR, that has the wrong sign, which is the reason why we had to use the antiparticle formulation 
\eC\ in \refeq{chiral}. The $1/\sqrt{3}$ normalization factor neatly goes into converting the single coupling constant 
$g$ of the model into the GWS model's two coupling constants $g$ and $g'\approx g/\sqrt{3}$.
	
	From the beginning it was noticed the graded group approach presented two serious 
difficulties: The first one was the so-called ``sign problem'', that arises from the fact that in non-abelian 
gauge (graded) theories the vector boson kinetic energy is constructed from the trace 
(supertrace) of the product of generators, that are (are not) positive-definite. A matrix 
normalization that is not positive-definite will give the wrong sign to the kinetic energy of some of 
the vector bosons.\citar{5} The second difficulty, that we are going to call the ``statistics problem'', had 
its origin in the fact that, if one is trying to reproduce the GWS model, Higgs fields have to be 
placed in the odd (or a-type) sector\citar{6} of the adjoint of \gsu, and therefore necessarily have to 
anticommute amongst themselves. On the other hand, they have spin zero and must therefore 
obey Bose-Einstein statistics, or problems with unitarity of the scattering matrix will ensue; thus, 
in contradiction with the graded group requirement, they have to commute among themselves. To 
overcome these difficulties was the motivation for much effort at the time.\citar{7} Many different ideas 
were tried such as adding extra odd dimensions to the spacetime manifold, and taking the 
Grassmann fields to be ghosts and not bosons, but, since neither difficulty could be resolved 
without causing worse difficulties, eventually interest on the subject waned.
 
	A few years ago a new branch of mathematics, noncommutative geometry,\citar{8} was 
developed. Its methods have been applied to the direct product of spacetime and spaces with a 
discrete number of points, with the result that some sectors of the standard model\citar{9} have been 
obtained. It has also been applied to other areas of theoretical physics.\citar{10} In the former case 
there appears a graded algebra of forms invariant under the algebra \gsuA. The GWS model 
with its \suu\ local gauge invariance appears if we define the lagrangian to be the trace 
(\letra{not} the supertrace) of group invariants. This way the sign problem is solved \letra{ab initio}. It is not 
clear to us just how successful this approach is in simplifying or unifying the Standard Model. 
Algebraic structures and ideas vary from paper to paper, sometimes even by the same author, so 
that, at this stage, we hesitate upon taking a position. 

	An attempt at generalizing a Yang-Mills theory using a graded gauge group will 
inevitably lead to the sign and the Higgs' statistics problems. Thus the question of attempting a 
generalization of the Yang-Mills covariant derivative using a non-graded Lie group arises 
naturally.\citar{11} A Higgs field assigned to the adjoint of the Lie group will now have the correct 
statistics since it would be even (that is, c-type). Furthermore, since Lie groups are invariant 
under traces, not supertraces, the kinetic energies of all the bosons will come out with the right 
sign. In this paper we show how to construct generalized Yang-Mills theories that are invariant 
under local gauge transformations of a Lie (not graded) group and use a covariant derivative with 
both scalar and vector fields. In the construction of these theories we also honor the condition, 
which we consider to be of an essential nature, that the lagrangians do not contain any 
differentiation operators acting indefinitely to the right. Terms of this type arise from powers of 
the covariant derivative, but we require that somehow they all cancel out. It goes without saying 
that such theories are also required to be Lorentz-invariant. It turns out that the only way to have 
a covariant derivative contain both vector and boson gauge fields is for it to transform as a 
4-vector contracted with the Dirac matrices, that is, as a slashed 4-vector.

	We apply these ideas to try to unify and reduce the number of different kinds of 
terms of the GWS model. If now one uses \su3\ as a gauge group, one \letra{almost} obtains the GWS 
model, the difference being that the hypercharge of the Higgs boson comes out wrong. An even 
more interesting result is that, if one uses the group \uu3\ and chooses a certain representation of 
the generators, the GWS comes out exactly, plus an additional scalar boson that is totally 
uncoupled both from all the other particles of the model. This unified theory has only two terms: 
the fermion kinetic energy, that is, the covariant derivative between two fermion fields, and the 
boson kinetic energy, that is, traces of powers of the covariant derivative, what is often called the 
curvature. We have not added a new level of symmetry breaking and in this sense we
have not performed a ``unification'' in the traditional sense: this theory has the
same number of Higgs bosons as the original GWS and they break the symmetry in
basically the same way. However, with the new covariant derivative the GWS can be
written in a simpler way in a larger group.

In this paper we do not go into the quark sector of the standard model. The presence
of two right quarks (as opposed to only one right lepton) per family results in a
different, more complicated situation, that we plan to address in a future. 

	In section 2 we show how to write a traditional abelian gauge theory with the 
covariant derivative transforming as the tensor product of two spinorial transformations. In 
section 3 we introduce the concept of scalar fields as gauge fields. In section 4 we present non-abelian 
Yang-Mills theories with mixed gauge fields. In section 5 we present two attempts at 
unification of the GWS model in the context of these ideas, using the Lie groups \su3 and \uu3. 
Finally, in section 6 we make a summary and conclude with a couple of remarks of what is wanting 
with the model.

\section{A vector field transforming under the tensor product of two spinorial representations}
\noindent
	In this section we are going to rewrite the quantum electrodynamics lagrangian
\ecuacion{
\lag_{QED} = \overline{\psi} i \scova \psi - \frac{1}{4} F^{\mu \nu} F_{\mu \nu}
\label{lagqed}
},
where $D_\mu \equiv \deriv{\mu} + ie A_\mu$ and $F_{\mu \nu} \equiv \deriv{\mu} A_\nu - \deriv{\nu} A_\mu = -ie^{-1} [D_\mu, D_\nu]$ using only contractions of 4-vectors 
with Dirac matrices, that is, avoiding dot products between the 4-vectors themselves. The reason 
for this is that there does not seem to be a way of generalizing gauge theories keeping the 
covariant derivative in the vector representation of the Lorentz group. One has to learn how to do 
everything with spinorial representations.

	The fermion field $\psi$ transforms under a local \uu1\ Lie group, so that, if $U=e^{-i\alpha(x)}$ 
is an element of this group, then $\psi \to U \psi$. To maintain gauge invariance it is necessary that the 
vector potential undergoes a gauge transformation, too, of the sort
\ecuacion{
A_\mu \to A_\mu + e^{-1} \deriv{\mu} \alpha
\label{transfgauge}
},

	From this transformation law and the definition of $D_\mu$, it is evident that the covariant derivative 
must transform as
\ecuacion{
D_\mu \to U D_\mu U^{-1}
\label{Dgauge}
}.

	We call a differentiation operator that acts only on immediately succeeding 
functions, but whose action then stops and does not differentiate any further functions to the 
right, \letra{a restrained operator}. Likewise, we call a differentiation operator \letra{unrestrained} if it keeps 
acting indefinitely to the right. As an example of the latter take the $\deriv{\mu}$'s in the covariant 
derivatives in \refeq{Dgauge}, that are acting to the right for we do not know how far. It is not admissible to 
leave unrestrained operators in a lagrangian because, first, what they can mean physically or 
mathematically is not clear, and, second, they are not gauge invariant. The boson kinetic energy in 
\refeq{lagqed} is proportional to $[D_\mu,D_\nu][D^\mu,D^\nu]$, each of whose factors is gauge invariant. On the other 
hand $D_\mu D_\nu$ is not gauge invariant as can be seen by direct calculation. The cause for this different 
behavior is evident from the equation
\ecuacion{
[D_\mu,D_\nu] f = \deriv{\mu} A_\nu f - A_\nu \deriv{\mu} f - \deriv{\nu} A_\mu f + A_\mu \deriv{\nu} f = (\deriv{\mu} A_\nu) f - (\deriv{\nu} A_\mu) f
\label{DmuDnu}
},
where it is seen how four unrestrained operators result in two restrained ones, thanks to Leibnitz' 
rule.

	If we represent by $S$ an element of the Lorentz spinor transformation group, so 
that $\psi \to S \psi$ under a Lorentz transformation, then, due to the homomorphism that exists between 
the vector and spinor representations of the Lorentz group, we have that $\sgauge \to S \sgauge S^{-1}$. This 
homomorphism allowed Dirac to write the equation of motion of electrons. But, is it possible to 
write the \letra{boson} kinetic energy with the vector potential transforming this same way? The 
following theorem answers this question in the affirmative:

\vspace*{12pt}
\noindent
{\bf Theorem:} Let $D_\mu = \deriv{\mu}+B_\mu$, where $B_\mu$ is a vector field. Then:
\ecuacion{
(\deriv{\mu} B_\nu - \deriv{\nu} B_\mu) (\partial^\mu B^\nu - \partial^\nu B^\mu) = \frac{1}{8} \tr^2 \scova^2 - \frac{1}{2} \tr \scova^4
\label{Teo}
},
where the trace is to be taken over the Dirac matrices. Notice the partials on the left of the 
equation are restrained, the ones on the right are not. To prove the Theorem we expand the 
covariant derivatives on the right side of \refeq{DmuDnu}, and then use the following trick, which makes the 
algebra manageable, in this and in more difficult examples in other sections. First, we define the 
differential operator $O \equiv \partial^2+2B \cdot \partial + B^2$. Notice that it does not contain any contractions with 
Dirac matrices, so that $\tr O = 4O$, $\tr O (\spart \sgaugeB) = 4O(\partial \cdot B)$, etc. It is not difficult to see then that 
$\scova^2 = O + (\spart \sgaugeB)$, where the slashed partial is acting \letra{only} on the succeeding slashed field. Now:
\ecuarreglo{
\frac{1}{8} \tr^2 [O + (\spart \sgaugeB)] - \frac{1}{2} \tr [O + (\spart \sgaugeB)]^2 &=& 2(\partial \cdot B)^2 - \frac{1}{2} \tr [(\spart \sgaugeB)(\spart \sgaugeB)] \nueva
&=& (\deriv{\mu} B_\nu - \deriv{\nu} B_\mu) (\partial^\mu B^\nu - \partial^\nu B^\mu)
\label{prueba}
}{}
and we have finished proving the Theorem. 

	With the aid of the Theorem we can rewrite the QED lagrangian in the form
\ecuacion{
\lag_{QED} = \overline{\psi} i \scova \psi + e^{-2} \iz( \frac{1}{32} \tr^2 \scova^2 - \frac{1}{8} \tr \scova^4 \de)
\label{lagQED}
},
whose Lorentz invariance can be proven using $\g{\mu} \to S \g{\mu} S^{-1}$ and the cyclic properties 
of the trace. We prove, as an example, the Lorentz invariance of $\scova^2$: $\tr \scova^2 \to \tr S \scova S^{-1} S \scova S^{-1} = \tr \scova^2$. 
Generally speaking, one could say that the cause for the term with 
the squared trace in \refeq{Teo} is the requirement that all differentiation operators be restrained.

\section{Abelian gauge theory using scalar boson fields}
\noindent
	We want to construct an analogue to quantum electrodynamics, but using scalar 
bosons as gauge fields. This is, of course, not the so-called scalar electrodynamics, where the 
fermion matter fields are substituted by scalar bosons. Here it is the gauge field itself that has 
become a real scalar boson $\higgs$. The transformation group is the same as in last section, and again 
to maintain gauge invariance \refeq{Dgauge} must hold. We now define the covariant derivative to be
\ecuacion{
D_\varphi = \spart - e \g5 \higgs
},
and we require the gauge field to transform as $\g5 \higgs \to \g5 \higgs - ie^{-1} \spart \alpha$. These conditions immediately 
assure us that transformation \refeq{Dgauge} holds in this case, too. There is an interesting point to be made 
here: if we now substitute this covariant derivative in the QED lagrangian \refeq{lagQED}, that was designed 
for \letra{vector} fields, we obtain the usual lagrangian for the interaction between a fermion and a real 
\letra{scalar} boson field. That is, if we \letra{assume} the lagrangian of the interaction to be
\ecuacion{
\lag_{S} = \overline{\psi} i D_\higgs \psi + e^{-2} \iz( \frac{1}{32} \tr^2 D_\higgs^2 - \frac{1}{8} \tr D_\higgs^4 \de)
\label{lagS}
},
then the expansion of the covariant derivative results in 
\ecuacion{
\lag_{S} = \overline{\psi} i \spart \psi - e \overline{\psi} i \g5 \higgs \psi + \frac{1}{2} (\deriv{\mu} \higgs)(\partial^\mu \higgs)
},
after a bit of algebra. This calculation is similar to the one done last section, but with the $\g5$ taking 
the place of the $\g{\mu}$'s of that previous calculation. The function of the $\g5$ is to ensure that the 
partial derivatives become restricted. If the gauge group had been non-abelian, then it would also 
ensure that we obtain commutators, not anticommutators, of all the boson fields.

\section{Non-abelian Yang-Mills theory with mixed gauge fields}
\noindent
	Consider a lagrangian that transforms under a non-abelian local Lie group that has 
$N$ generators. The fermion or matter sector of the non-abelian lagrangian has the form $\overline{\psi} i \scova \psi$, 
where $D_\mu$ is a covariant derivative chosen to maintain gauge invariance. This term is invariant 
under the transformation $\psi \to U \psi$, where $U=U(x)$ is an element of the fundamental 
representation of the group. The covariant derivative is $D_\mu = \deriv{\mu} + A_\mu$, where $A_\mu = ig \Amu (x) T^a$ is 
an element of the Lie algebra and $g$ is a coupling constant. We are assuming here that the set of 
matrices $\{ T^a \}$ is a representation of the group's generators. Gauge invariance of the matter term 
is assured if
\ecuacion{
A_\mu \to U A_\mu U^{-1} - (\deriv{\mu} U) U^{-1}
},
or, what is the same,
\ecuacion{
\sgauge \to U \sgauge U^{-1} - (\spart U) U^{-1}
}.

	We have already seen how scalar fields can function as gauge fields. Our aim in 
this section is to construct a non-abelian theory that uses both scalar and vector gauge fields. We 
proceed as follows. For every generator in the Lie group we choose one gauge field, it does not 
matter whether vector or scalar. As an example, suppose there are $N$ generators in the Lie group; 
we choose the first $N_V$ to be associated with an equal number of vector gauge fields and the last 
$N_S$ to be associated with an equal number of scalar gauge fields. Naturally $N_V + N_S = N$. Now we 
construct a covariant derivative $D$ by taking each one of the generators and multiplying it by one 
of its associated gauge fields and summing them together. The result is
\ecuacion{
D \equiv \spart + \sgauge + \Phi
\label{general}
},
\ecuarreglo{
\sgauge & \equiv & \g{\mu} A_\mu \equiv ig \g{\mu} \Amu T^a \coma, \espacio a = 1, \ldots, N_V \ptc,
\Phi    & \equiv & \g5 \higgs \equiv -g \g5 \higgs^b T^b \coma, \espacio b = N_V+1, \ldots, N  \nonumber
}.

Notice the difference between $A_\mu$ and $\Amu$, and between $\higgs$ and $\higgs^b$. We take the gauge 
transformation for these fields to be
\ecuacion{
\sgauge + \Phi \to U (\sgauge + \Phi) U^{-1} - (\spart U) U^{-1}
},
from which one can conclude that
\ecuacion{
D \to U D U^{-1}
}.

The following lagrangian is constructed based on the requirements that it contains only matter 
fields and covariant derivatives, and that it possesses both Lorentz and gauge invariance:
\ecuacion{
\lag_{NA} = \overline{\psi} i D \psi + \frac{1}{2g^2} \widetilde{\tr} \iz( \frac{1}{8} \tr^2 D^2 - \frac{1}{2} \tr D^4 \de)
\label{lagNA}
},
where the trace with the tilde is over the Lie group matrices and the one without it is over 
matrices of the spinorial representation of the Lorentz group. The additional factor of $1/2$ that the 
traces of \refeq{lagNA} have with respect to \refeq{Teo} comes from normalization \refeq{norma}, that is the usual 
one in the non-abelian case.

	Although lagrangian $\lag_{NA}$ was constructed based only on the requirements just 
mentioned, it is an interesting fact that the expansion of the covariant derivative into its 
component fields results in expressions that are traditional in Yang-Mills theories. The reader who 
wishes to make the expansion herself can substitute \refeq{general} in \refeq{lagNA}, keeping in mind the derivatives 
are unrestrained, and aim first for the intermediate result
\ecuarreglo{
\frac{1}{16} \tr^2 D^2 - \frac{1}{4} \tr D^4 &=& \iz( (\partial \cdot A)+A^2 \de)^2 - \frac{1}{4} \tr \iz( (\spart \sgauge)+\sgauge \sgauge \de)^2 \nueva
   & & - \frac{1}{4} \tr \iz( (\spart \Phi)+\{ \sgauge \Phi \} \de)^2
\label{expandido}
},
where the curly brackets denote an anticommutator. (We recommend to use here the same trick 
explained in section 2.) Notice in this expression that the differentiation operators are restrained, 
and that the two different types of gauge fields appear in an anticommutator. One of the effects of 
the $\g5$ in \refeq{general} is to turn this anticommutator into a commutator through the use of the properties 
of Clifford algebras. Substituting \refeq{general} in \refeq{expandido} and in the matter term of \refeq{lagNA} we obtain the 
non-abelian lagrangian in expanded form:
\ecuarreglo{
\lag_{NA} &=& \overline{\psi} i(\spart + \sgauge) \psi - g \overline{\psi} i \g5 \higgs^b T^b \psi + \frac{1}{2 g^2} \widetilde{\tr} \iz( \deriv{\mu} A_\nu - \deriv{\nu} A_\mu + [A_\mu, A_\nu] \de)^2 \nueva
   & & + \frac{1}{g^2} \widetilde{\tr} \iz( \deriv{\mu} \higgs + [A_{\mu}, \higgs] \de)^2
\label{lagexpandido}
}.

The reader will recognize familiar structures: the first term on the right looks like the usual matter 
term of a gauge theory, the second like a Yukawa term, the third like the kinetic energy of vector 
bosons in a Yang-Mills theory and the fourth like the gauge-invariant kinetic energy of scalar 
bosons in the non-abelian adjoint representation. It is also interesting to observe that, if in the last 
term we set the vector bosons equal to zero, then the term simply becomes $\sum_{b,\mu} \frac{1}{2} \deriv{\mu}\higgs^b \partial^\mu \higgs^b$, the 
kinetic energy of the scalar bosons. We have constructed a generic non-abelian gauge theory with 
gauge fields that can be either scalar or vector.

\section{The GWS model using \bfsu3\ and \bfuu3}
\noindent
	The obvious choice for a small group to simplify and unify the GWS model using a 
generalized gauge theory is \su3, because it has the correct hypercharge numbers for the leptons 
of a chiral multiplet. Furthermore, it has 8 generators, while the GWS model has precisely 8 
bosons: 4 vector ones and the 4 Higgs real scalar fields. Unfortunately, this choice does not work 
as we shall soon see. Let $\Amu$, $a=1,2,3$, and $B_\mu$ be four vector fields to which we associate the 
four Gell-Mann generators $\ld{a}$, $a=1,2,3$, and $\ld{8}$, respectively. Let $\higgs^b$, $b=4,5,6,7$, be four real 
scalar fields, to which we associate the remaining Gell-Mann generators $\ld{b}$, $b=4,5,6,7$. The 
covariant derivative can be found following the prescription given in \refeq{general}. Using the usual 
representation of the Gell-Mann matrices and $g$ as coupling constant it can be explicitly written 
as follows:
\ecuacion{
D = \spart + \frac{g}{2} \pmatrix{i \bfsgauge \cdot \bfsigma + i \sgaugeB \uno / \sqrt{3} & -\sqrt{2} \g5 \widehat{\higgs} \cr
-\sqrt{2} {\widehat{\higgs}}^{\dag} \g5 & -2i \sgaugeB / \sqrt{3} }
\label{DSU3}
},
where the $\sigma^a$, $a=1,2,3$, are the Pauli matrices, $\uno$ is a $2 \times 2$ matrix, and
\ecuacion{
\widehat{\higgs} = \frac{1}{\sqrt{2}} \pmatrix{ \higgs^4 - i \higgs^5 \cr \higgs^6 - i \higgs^7 }
}.

The lagrangian of the model is then precisely $\lag_{NA}$, as expressed by \refeq{lagNA}, with the covariant 
derivative given by \refeq{DSU3} and the fermion triplet by $\psi_L$, the chiral triplet of equation \refeq{chiral}. To 
obtain the expanded form of the boson kinetic energy sector through straightforward calculation 
is very messy, but through the use of formula \refeq{lagexpandido} it is possible to arrive at the following 
expression without much trouble:
\ecuarreglo{
\lag_{NA} &=&\overline{\theta_L} (i\spart - \frac{1}{2} g \sgauge^a \sigma^a - \frac{1}{2} g' \sgaugeB) \theta_L + \overline{e_R} (i \spart - g' \sgaugeB) e_R \nueva
          & &+i \frac{g}{\sqrt{2}} \overline{e_L^c} \widehat{\higgs}^{\dag} \theta_L +i \frac{g}{\sqrt{2}} \overline{\theta_L} \widehat{\higgs} e_L^c \nueva
          & &-\frac{1}{4} {\bf A_{\mu \nu} \cdot A^{\mu \nu}} -\frac{1}{4} B_{\mu \nu} B^{\mu \nu} \nueva
          & &+\iz| (\deriv{\mu} +\frac{1}{2} ig \Amu \sigma^a + \frac{3}{2} ig' B_\mu) \widehat{\higgs} \de|^2
\label{lagSU3}
},
where $\theta_L = (\nu_L, e_L)^T$ and we have introduced the symbol $g'\equiv g/\sqrt{3}$. This is \letra{almost} the lagrangian 
of the GWS model for a Weinberg angle $\theta_W = 30^\circ$ (a value very close to the experimental one), 
except that the hypercharge of the Higgs comes out as $3$ and not $-1$ as it should be. This detail 
dooms the model. Notice too the Yukawa terms are actually null, as one can see from chirality 
considerations.

	Let us look at the problem of the hypercharge in more detail. Let
\ecuacion{
\ld{Y} = \frac{1}{\sqrt{3}} \diag (x,x,y)
}{}
be some generator associated with the vector field $B_\mu$, the one that eventually becomes the 
isospin singlet in the GWS model; then the hypercharge of the Higgs is $Y=x-y$. In the example 
above we were using $\ld{8}$, shown in \refeq{gene8}, so that $Y=1-(-2)=3$. If instead of \su3\ we had been 
using \gsu\ then the choice for $\ld{Y}$ would have been $\ld{0}$, shown in \refeq{gene0}, and $Y=1-(+2)=-1$, the 
correct value. So it seems we have reached an \letra{impasse}: the correct hypercharge is given precisely 
by the group we do not want to use. The way out of it is to realize that a very similar group, 
\uu3, has a representation where $\ld{0}$ appears. In other words, it is not necessary to go to graded 
groups to obtain the correct hypercharge. We will see now how this comes about.

	A representation of \uu3\ is, for instance, the 8 Gell-Mann generators $\ld{a}$, $a=1,\ldots ,8$ plus 
another (not traceless) $3 \times 3$ matrix which we take to be the normalized unit matrix:
\ecuacion{
\ld{9} = \sqrt{\frac{2}{3}} \diag (1,1,1)
}.

We now define a new matrix
\ecuacion{
\ld{10} = \sqrt{\frac{2}{3}} \diag (1,1,-1)
},
and make the observation that it and $\ld{0}$ can be expressed as linear combinations of $\ld{8}$ and $\ld{9}$:
\ecuarreglo{
\ld{10} &=& \frac{2 \sqrt{2}}{3} \ld{8} + \frac{1}{3} \ld{9} \ptc,
\ld{0} &=& -\frac{1}{3} \ld{8} + \frac{2 \sqrt{2}}{3} \ld{9}
}.
Notice that $\ld{10}$ and $\ld{0}$ are orthonormal under \refeq{norma}. The set of 9 matrices $\ld{10}$, $\ld{0}$ and $\ld{a}$, 
$a=1,\ldots ,7$ forms a representation of \uu3, since each matrix is a linear combination of the 
generators of another representation. So we have found $\ld{0}$ in \uu3, and associating it with $B_\mu$ we 
make sure that the hypercharge for the Higgs comes out correctly.

	This new group has 9 generators, and, therefore, 9 bosons. This is cause for concern, since 
the extra boson could upset the precise clockwork phenomenology of the GWS model. We have 
in principle the option of taking it to be either scalar or vector, but the second option would 
present us with an unphysical extra vector boson. On the other hand, if we take it to be a scalar 
boson, it is an interesting fact that the boson completely decouples from the rest of the particles 
and we obtain again lagrangian \refeq{lagSU3}, but with the correct hypercharge for the Higgs bosons. Now 
$\psi$ is given by \refeq{nonchiral} and not by \refeq{chiral}, as in the \su3\ case.

	To understand why the new scalar boson, that we shall call $\Upsilon$, decouples, go back to \refeq{lagexpandido}, 
the non-abelian lagrangian, and notice that the scalar bosons only appear in the second and fourth 
terms of the right of the equation. Since the scalar boson is associated with $\ld{10}$, which is a 
diagonal generator, in the second term the spinorial degrees of freedom are all multiplying their 
respective conjugates, that have the opposite chirality, and therefore all products are null. In the 
fourth term within the parenthesis there appears, for the case of the new scalar, the expression $[A_\mu, \ld{10} \Upsilon]$,
that is, the commutator of a block diagonal matrix and $\ld{10}$, and, therefore, zero. The 
only term that remains in the lagrangian that contains $\Upsilon$ is its kinetic energy. It would seem this 
field is massless, as it does not couple with the Higgs.

	We again expand $\lag_{NA}$, this time using the new set of generators and the nonchiral 
fermion triplet $\psi$ of equation \refeq{nonchiral}, with the result:
\ecuarreglo{
\lag_{NA} &=&\overline{\theta_L} (i\spart - \frac{1}{2} g \sgauge^a \sigma^a - \frac{1}{2} g' \sgaugeB) \theta_L + \overline{e_R} (i\spart - g' \sgaugeB) e_R \cr
          & &+i \frac{g}{\sqrt{2}} \overline{e_R} \widehat{\higgs}^{\dag} \theta_L -i \frac{g}{\sqrt{2}} \overline{\theta_L} \widehat{\higgs} e_R \cr
          & &-\frac{1}{4} {\bf A_{\mu \nu} \cdot A^{\mu \nu}} -\frac{1}{4} B_{\mu \nu} B^{\mu \nu} \cr
          & &+\iz| (\deriv{\mu} +\frac{1}{2} ig \Amu \sigma^a - \frac{1}{2} ig' B_\mu) \widehat{\higgs} \de|^2 + \frac{1}{2} (\deriv{\mu} \Upsilon)(\partial^\mu \Upsilon)
\label{lagU3}
},
which is the lagrangian of the GWS model. Notice the two Yukawa terms do not vanish in this 
lagrangian, as they did in \refeq{lagSU3}. The difference of sign between them does not matter, as it simply 
depends on what phase of the \eR\ we decide to use.
 
	In Yang-Mills theories when the covariant derivative acts on a field $X$, its gauge fields 
acquire certain coefficients called the ``charges'' of each gauge field with respect to $X$. In this 
generalized gauge theory the same thing has to be done. Thus, when $D$ acts on the leptonic 
triplet, its gauge fields are going to be multiplied by constants we shall call ``leptonic charges'' $Q_V$ 
and $Q_S$, with the result $D \psi = (\spart + Q_V \sgauge + Q_S \Phi) \psi$. From our knowledge of the standard model we 
conclude that $Q_V=1$, and that there are three $Q_S$, one for each generation, and have rather small 
values. Lagrangian \refeq{lagU3} is actually for one generation only, and it should include its respective $Q_S$ 
as a coefficient in the two Yukawa terms. We have no \letra{a priori} knowledge of the values of the 
leptonic charges.

\section{Summary and remarks}
\noindent
	We found a particular way of expressing the kinetic energy of vector bosons (what in 
principal vector bundles is called the curvature) so that those fields appear contracted only with 
Dirac matrices. From here we were able to generalize the concept of covariant derivative, so that 
it included both scalar and vector bosons. Using this generalized derivative one can write the 
GWS model using only two terms (a curvature and a matter term), and unify the two groups (and 
their coupling constants) into one. This derivative has to transform as a slashed 4-vector.

	The unification group is not \su3, that gives the wrong value for the hypercharge of the 
Higgs field, but \uu3\ instead, that predicts correctly all the quantum numbers of the GWS model. 
This is possible because there is a representation of this group that contains the same generator of 
\gsu\ that gives the right hypercharge to the Higgs bosons. Since we did not use a graded 
group, the Higgs fields obey, correctly, Bose-Einstein statistics, and the kinetic energy terms of 
\letra{all} vector bosons have the right sign.	An extra scalar boson has to included (since \uu3\ has one 
more generator than \su3) but it automatically decouples from the rest of the model becoming 
an unobservable particle. It is probably massless, since it does not couple with the Higgs field, 
either.

In the GWS model symmetry breaking is achieved spontaneously through the
introduction of a potential of the form $V(\higgs)$. In the model presented here 
this potential also has to be introduced explicitly, as it does not appear in its
sole
two terms, kinetic energy of bosons and of fermions. In this it differs from the
noncommutative geometry results for the standard model, that implicitly include
this potential. The terms for the Higgs potential appear in our model, but they 
cancel since we introduce counterterms for the purpose of avoiding the presence of
differentiation operators acting indefinitely to the right. (See equation
\refeq{Teo}.) These counterterms also avoid other unwanted self-interacting terms.

The Yukawa coupling constants appear in a natural way as generalized gauge charges
of the lepton triplet $\psi$, but the model sheds no light about what those values
may be, or why there is a different value for each generation.

\nonumsection{References}

\end{document}